\documentclass[aps,showpacs,prd,onecolumn,amsmath,amssymb]{revtex4}
\usepackage{color}
\usepackage{graphics}
\usepackage{epsfig}
\usepackage{amsmath}
\usepackage{amssymb}
\usepackage{amsthm}
\usepackage[mathscr]{eucal}
\newcommand{\slsh}[1]{{\not \! #1}}
\newcommand{\Eq}[1]{{Eq.~({\ref{#1}})}}

\newcommand{\bea}{\begin{eqnarray}}
\newcommand{\eea}{\end{eqnarray}}
\newcommand{\beas}{\begin{eqnarray*}}
\newcommand{\eeas}{\end{eqnarray*}}
\newcommand{\sumint}{\sum\!\!\!\!\!\!\!\!\int}

\begin{document}


\title{Non-perturbative Euler-Heisenberg Lagrangian and Paraelectricity in Magnetized Massless QED}
\author{Efrain J. Ferrer, Vivian de la Incera and Angel Sanchez}
\affiliation{ Department of Physics, University of Texas at El Paso,
  500 W. University Ave., El Paso, TX 79968, USA}

\begin{abstract}
In this paper we calculate the non-perturbative Euler-Heisenberg Lagrangian for massless
QED in a strong magnetic field $H$, where the breaking of the chiral symmetry is dynamically catalyzed by the external magnetic field via the formation of an electro-positron condensate. This  chiral condensate leads  to the generation of dynamical parameters that have to be found as  solutions of non-perturbative Schwinger-Dyson equations. Since the electron-positron pairing mechanism leading to the breaking of the chiral symmetry is mainly dominated by the contributions from the infrared region of momenta much smaller than $\sqrt{eH}$, the magnetic field introduces a dynamical ultraviolet cutoff in the theory that also enters in the non-perturbative Euler-Heisenberg action. Using this action, we show that the
system exhibits a significant paraelectricity in the direction parallel to the magnetic field. The nonperturbative nature of this effect is
reflected in the non-analytic dependence  of the obtained electric susceptibility on the
fine-structure constant. The strong paraelectricity in the field direction is linked to the orientation of the electric dipole moments of the pairs that form the chiral condensate. The large electric susceptibility can be
used to detect the realization of the magnetic catalysis of
chiral symmetry breaking in physical systems.
\end{abstract}
\maketitle

\section{Introduction}

The study of relativistic quantum theories in the presence of strong magnetic fields has been an active topic of research in the physics literature over many years \cite{QED-B}.  The observation of magnetars' surface magnetic fields in the range of $10^{12}G -10^{15}$G  \cite{Magnetars}, together with estimates of  inner magnetic fields between $10^{18}$G and $10^{20}$G,  for star cores made of nuclear \cite{NS} and quark matter \cite{QS} respectively, have contributed to renovate the interest in this area. In addition, the production of very strong magnetic fields $\sim 2m^2_\pi
(\sim10^{18} G)$ in non-central heavy-ion collisions at RHIC \cite{Fukureview}, and the prediction of even larger values, $eH\sim 15 m^2_\pi
(\sim 10^{19} G)$, at future LHC experiments \cite{LHC}, has also served to inspire new studies on the effects of strong magnetic fields in QED and QCD.

On the other hand,  the effects of strong magnetic fields in relativistic theories is also important for planar condensed matter. The low-energy excitation quasiparticle spectra in systems like the pyrolitic graphites (HOPG) \cite{Semenoff, graphite} and graphene \cite{graphene} are characterized by a linear
dispersion and then the dynamics of their charge carriers can be described by a
"relativistic" quantum field theory of massless fermions in (2+1) dimensions ~\cite{Semenoff, Wallace}.  Because the properties of these systems can be modified in the presence of a strong magnetic field \cite{grapheneBexp},  many works \cite{MC-Applications, grapheneBtheo} have been aimed at understanding the mechanisms  behind the observed behaviors.

Massless QED in the presence of a magnetic field exhibits a peculiar
phenomenology. Due to the Landau quantization of the fermion's
transverse momentum in a magnetic field, the dynamics of the particles in the lowest
Landau level (LLL) is (1+1)-dimensional. This dimensional
reduction favors the formation of a chiral condensate, even at the
weakest attractive coupling, because there is no energy gap between the infrared
fermions in the LLL and the antiparticles in the Dirac sea. This
phenomenon is known as the magnetic catalysis of chiral symmetry
breaking (MC$\chi$SB).  The MC$\chi$SB modifies the vacuum properties
and induces dynamical parameters that depend on the applied
field. This effect has been actively investigated for the last two
decades~\cite{severalaspects}-\cite{prlFIS}. In the original
studies of the MC$\chi$SB ~\cite{severalaspects}-\cite{leungwang}, the
catalyzed chiral condensate was assumed to give rise only to a
dynamical fermion mass. Recently, however, it has become clear
\cite{ferrerincera} that besides the dynamically generated mass, the
MC$\chi$SB inevitably produces also a dynamical anomalous magnetic
moment (AMM), because this second parameter does not break any
symmetry that has not already been broken by the chiral condensate and
the magnetic field. The dynamical AMM leads, in turn, to a
non-perturbative Lande g-factor and Bohr magneton proportional to the
inverse of the dynamical mass. The induction of the AMM leads to a
non-perturbative Zeeman effect \cite{ferrerincera}. An important aspect of the MC$\chi$SB is its universal character. It
will occur in any relativistic theory of interactive massless fermions
in a magnetic field. The MC$\chi$SB has been proposed as the mechanism
explaining various effects in quasiplanar condensed matter
systems~\cite{MC-Applications}.

A drawback of the MC$\chi$SB phenomenon is that the dynamical
parameters (mass and AMM) are extremely small even at relatively high
fields. Hence, it would be important to have an
independent way to experimentally detect this phenomenon. In a recent letter \cite{prlFIS}, we found that by measuring the induced electric polarization of the magnetized medium, compelling
evidence in favor or against the existence of MC$\chi$SB can be obtained.

In this paper we find the non-perturbative Euler-Heisenberg Lagrangian of massless QED in a strong magnetic field, a theory in which the chiral symmetry is  broken via the MC$\chi$SB mechanism and both the fermion mass and the AMM are dynamically generated \cite{ferrerincera}. As it is known, the Euler-Heisenberg Lagrangian encodes the vacuum structure of
QED in the presence of a constant
electromagnetic field. This Lagrangian, first found by Euler and
Heisenberg~\cite{Euler} and then reformulated by Schwinger~\cite{Schwinger}, is a
low-energy effective Lagrangian describing the non-linear dynamics of the electromagnetic field. It is obtained from a single electron loop
coupled to the external electromagnetic field. This formulation has been widely
used to describe a variety of electromagnetic phenomena like photon
splitting~\cite{Adler}, light scattering~\cite{Eber},
birefringence~\cite{Klein},  pair production~\cite{Ruffini}, etc.  However, to the best of our knowledge, the low-energy Euler-Heisenberg Lagrangian has never been derived for a theory with dynamically generated parameters, which is the goal of this work.  In the present case, the electron propagator in the loop will be the full propagator taken in the ladder approximation and in the presence of a strong magnetic field. Hence, our effective Euler-Heisenberg Lagrangian will be essentially nonperturbative because the electron self-energy that will be used to obtain it is found by a consistent resummation of an infinite number of rainbow diagrams. The nonperturbative origin of the effective action will be instrumental to show the paraelectric behavior of the strongly magnetized massless QED.

The paper is organized as follows. In
Sec.~\ref{MCSB}, we present a brief review of the non-perturbative
MC$\chi$SB phenomenon in massless QED.  The dynamical parameters of the theory are found as solutions of the Schwinger-Dyson equations (SDE) in the
ladder approximation. These solutions will be later used to find the electric response of the magnetized medium to a weak electric field that is used as a probe. In Sec.\ref{EulerHeisenberSec}, we employ
Ritus's eigenfunctions for constant and parallel electric and magnetic fields and the
path integral formulation of Euclidean QED with a full fermion propagator, to obtain the non-perturbative Euler-Heisenberg action in a strong magnetic field. In Sec.~\ref{paraelecsec},
we use the non-perturbative Euler-Heisenberg action to derive the
electric susceptibility of the magnetized medium at strong magnetic field. Direct applications of the outcome of this paper to condense
matter systems, like graphene, are discussed in
Sec.~\ref{conclusion}. In Appendix~\ref{ritusappendix},  the $\mathbb{E}_{p}$ functions are obtained in Euclidean space for the case of parallel electric and magnetic fields. Finally, in Appendix B, we present an alternative approach, which starts from the effective (1+1)-dimensional theory of the electrons in the LLL, to obtain the Euler-Heisenberg Lagrangian of a strongly magnetized system.

\section{M$\chi$SB in Massless QED reviewed}~\label{MCSB}

In this section, for the sake of completeness and understanding, we review the realization of the M$\chi$SB in massless QED, taking into account the dynamical generation of an AMM. In this approach, the electron Green's function depends
non-perturbatively on the magnetic field, as well as on the dynamical mass, $M^l,$ and the induced AMM, $T^l$, for each Landau level $l$. A more detailed description can be found in
Ref.~\cite{ferrerincera}.

The Green's function that describes the motion of an electron in an
arbitrary electromagnetic field satisfies in coordinate space the equation
\bea
\left[\gamma_\mu\Pi^\mu-\Sigma(x,y)\right]G(x,y)=\delta^4(x-y)
\label{eqfull}
\eea
where $\Pi_{\mu}=i(\partial_\mu+ieA_\mu)$ and $\Sigma(x,y)$ are the
covariant derivative depending on the external field and the electron self-energy, respectively.

In a constant magnetic field the fermion self-energy operator is a combination
of the operators $\gamma_\mu\Pi^\mu$, $\sigma_{\mu\nu}F^{\mu\nu}$,
$(F^{\mu\nu}\Pi_\mu)^2$ and $(\gamma_\mu\Pi^\mu)^2$~\cite{Ritus:1978cj,WT}.
As all these operators commute with the later, the fermion self-energy
operator  will be diagonal in the basis spanned by the matrix-eigenfunctions $\mathbb{E}_p$
of $(\gamma_\mu\Pi^\mu)^2$,
\bea
   (\gamma\cdot\Pi)^2\mathbb{E}_p=\overline{p}^2\mathbb{E}_p\, .
\label{ad1}
\eea
For a magnetic field along the $x_3$-direction, the eigenvalue $\overline{p}_\mu$, in the Landau-like gauge $A_\mu=(0,0,Hx_1,0)$,
is given by
  $\overline{p}_\mu=(p_{0},0,sgn(eH)\sqrt{2|eH|l},p_{3})$ with
$l=0,1,2,...$ denoting the Landau Levels. From now on, we will consider in this section that $sgn(eH)>0$.

From the physical point of view, the $\mathbb{E}_p$ functions are the eigenfunctions of the asymptotic states of the charged particle in the
presence of a constant magnetic field. The use of these eigenfunctions to
diagonalize the self-energy operator in momentum space was originally done by Ritus \cite{Ritus:1978cj} for the case of fermions. His method was later
extended to charged spin-one fields in \cite{efi-ext}.

In the chiral representation, the  $\mathbb{E}_p$ functions take the form
\bea
\mathbb{E}_p=\sum_{\sigma=\pm1}E_{p\sigma}(x)\Delta(\sigma),
\label{ad2}
\eea
where $\Delta(\sigma)=(I+i\sigma \gamma^1\gamma^2)/2$
is the spin projector with $\sigma=\pm1$. The eigenfunctions
$E_{p\sigma}(x)$ are given by
\bea
   E_{p\sigma}(x)=N_n e^{-i(p_0x^0+p_2x^2+p_3x^3)}D_n(\rho)\,,
\label{ad4}
\eea
with $D_n(\rho)$ the parabolic cylinder functions of argument
$\rho=\sqrt{2|eH|}(p_2/eH+x^1)$ and index
$n=n(l,\sigma)\equiv l+\frac{\sigma-1}{2}${, and
  normalization constant $N_n=(4\pi eH)^{1/4}/\sqrt{n!}$.}

The $\mathbb{E}_p$'s are orthonormal \cite{leungwang}
\bea
    \int d^4x \overline{\mathbb{E}}_p(x) \mathbb{E}_{p'}(x)
     =(2\pi)^4 \hat{\delta}^{(4)}(p-p')\Pi(l)
\label{ad6}
\eea
and complete
\bea
   \sumint \frac{d^4p'}{(2\pi)^4}
    \mathbb{E}_{p'}(x)\overline{\mathbb{E}}_{p'}(y)=\delta^4(x-y),
\label{completenessrelation}
\eea
with notation $\overline{\mathbb{E}}_p(x)=\gamma^0\mathbb{E}_p^\dagger\gamma^0$,
and
\bea
  \hat{\delta}^{(4)}(p-p')
   =\delta^{ll'}\delta(p_0-p'_0)\delta(p_1-p'_1)\delta(p_3-p'_3),
\label{ad7}
\eea

\bea
\Pi(l)=\Delta(sgn(eH))+\Delta(-sgn(eH))(1-\delta_{0l}),
\label{ad7-1}
\eea
where $\Pi(l)$ takes into account that for the lowest Landau level only one
spin projection is allowed. Since we consider $eH>0$,  the spin projection at the LLL will be associated with $\Delta(+)$.

The $\mathbb{E}_p$ functions satisfy the relations
\bea
   (\gamma\cdot\Pi)\mathbb{E}_p=\mathbb{E}_p(\gamma\cdot\bar{p})
\label{ad8}
\eea
and
\bea
   (Z_{||}\slsh{\Pi}_{||}+Z_\perp \slsh{\Pi}_\perp)\mathbb{E}_p(x)
    =\mathbb{E}_p(Z_{||}\slsh{\bar{p}}_{||}+Z_\perp\slsh{\bar{p}}_\perp)
\label{ad9}
\eea
where $\bar{p}^{||}_\mu=(p_0,0,0,p_3)$ and
$\bar{p}^\perp_\mu=(0,0,\sqrt{2eH l},0)$, and, $Z_{||}$ and
$Z_{\perp}$ are the longitudinal and transverse wave-function renormalization coefficients, respectively.

Transforming to momentum space with the help of the $\mathbb{E}_p$ functions, it is easy to check, using Eqs.~(\ref{ad6})-(\ref{ad9}), that the fermion self-energy operator becomes diagonal in momentum
space
\bea
  \Sigma(p,p')&\equiv&
        \int d^4xd^4y\overline{\mathbb{E}}_p(x)\Sigma(x,y)\mathbb{E}_{p'}(y)
    =(2\pi)^4\hat{\delta}^{(4)}(p-p')\Pi(l)\widetilde{\Sigma}^l(\overline{p})
 \label{ad10}
\eea
with
\bea
    \widetilde{\Sigma}^l(\overline{p})
     &=&Z_{||}^l\slsh{\bar{p}}_{||}+Z^l_{\perp}\slsh{\bar{p}}_\perp
      +(M^l+T^l)\Delta(+)
    +(M^l-T^l)\Delta(-)
\label{ad11}
\eea
where $M^l$ and  $T^l$ are  the fermion dynamical mass and AMM for each LL, respectively. They have to be found self-consistently by
solving the non-perturbative Schwinger-Dyson equations (SDE) of  the theory. The
presence of $\Pi(l)$ in \Eq{ad10} ensures the separation of
the LLL from the rest of the levels due to its lack of spin degeneracy.

Applying the $\mathbb{E}_p$ transformation in (\ref{eqfull}), the full
propagator in momentum space can be found as
\begin{equation}\label{full-FP}
   G^{l}(p,p')\equiv
        \int d^4xd^4y\overline{\mathbb{E}}_p(x)G(x,y)\mathbb{E}_{p'}(y)=(2\pi)^4\widehat{\delta}^{(4)}(p-p')
           \Pi(l)\widetilde{G}^{l}(\overline{p})
\end{equation}
where
\bea
  \tilde{G}^l(\overline{p})&=&
    \sum_{\sigma,\overline{\sigma}=\pm 1}
    \frac{N^l(\sigma T,\overline{\sigma}V_{||})
          -iV_\perp^l(\Lambda_\perp^+-\Lambda^-_\perp)}
         {D^l(\sigma\overline{\sigma}T)}
    \Delta(\sigma)\Lambda^{\overline{\sigma}}_{||}
\nonumber \\
\label{ad12}
\eea
with
\bea
   \Lambda^{\overline{\sigma}}_{||}&=&
         \frac{1}{2}
         \left(1+{\overline{\sigma}}\frac{\slsh{\overline{p}}_{||}}{|\overline{p}_{||}|}
         \right)
\nonumber \\
   \Lambda^\sigma_{\perp}&=&
         \frac{1}{2}
         \left(1+i\sigma\gamma^2
         \right)
\nonumber \\
   N^l(\sigma T,\overline{\sigma}V_{||})
     &=&\sigma T^l-M^l-\overline{\sigma}V_{||}^l
\nonumber \\
   D^l(\sigma {\overline{\sigma}}T)&=&(M^l)^2-(V_{||}^l-\sigma{\overline{\sigma}} T^l)^2+(V_\perp^l)^2
\nonumber \\
    V_{||}^l&=&(1-Z_{||}^l)|\overline{p}_{||}|
\nonumber \\
    V_{\perp}^l&=&(1-Z_\perp^l)|\overline{p}_{\perp}|.
\label{ad13}
\eea

The non-perturbative SDE for the self-energy can be solved using the ladder approximation, where it takes the form
\bea
   \Sigma(x,y)=ie_{R}^2\gamma^\mu G(x,y)\gamma^\nu D_{\mu\nu}(x-y).
\label{auto1}
\eea
Here, the full vertex is replaced by the free one, $D_{\mu\nu}(x-y)$ is the free photon propagator in
the Feynman gauge,  $G(x,y)$ is the full fermion propagator
(\ref{ad12}), and $e_{R}$ is the renormalized coupling constant at the scale $\sqrt{eH}$.  In general, the truncation used for the resummation of the diagrams contributing to the SDEs in a given nonperturbative approximation can lead to gauge-dependent results. Therefore, care should be taken in choosing a gauge where the considered truncation is consistent and the physical parameters like the dynamical mass, the renormalized coupling, etc., are all gauge-independent. This is a well-known and old problem. For example, for the approximation used in the Abelian theory of fermions and vector mesons discussed in Ref. \cite{CJT-PRD10-74}, the consistent gauge was the Landau gauge. In this special gauge all the vacuum diagrams with the Goldstone pole typical of theories with spontaneous symmetry breaking cancelled out, and as a consequence, the full vertex was replaced by the free one, so the truncation used there became reliable. In the presence of an external magnetic field, the consistent gauge in the ladder approximation is the Feynman gauge \cite{leeleungng}. One can show that in this gauge all the terms coming from loop corrections to the vertex cancel out and the full vertex reduces to the free one.  The gauge-independence of the physical parameters in this case was verified in \cite{WT}  by showing that in this special gauge the solution of the ladder SDE satisfies the Ward-Takahashi identities.  Given that in the ladder approximation both the vertex and the photon propagator are replaced by the bare ones, the ultraviolet divergences that would appear in a perturbative approach all cancel out, as long as the consistent gauge is used, so the photon field renormalization constant $Z_3$, which also enters in the coupling constant renormalization, reduces to 1, leading to renormalized coupling and photon field at the characteristic scale of the infrared region where the SDE is solved, i.e.  $\sqrt{eH}$ \cite{NPB462}. Of course, any change in the approximation means taking a different truncation of the infinite series of diagrams and hence requires to find a new consistent gauge where the physical parameters result gauge-independent again. For example, going beyond the ladder approximation in the presence of a magnetic field, as in the so-called improved rainbow approximation that includes the polarization effects in the photon propagator, requires to use a non-covariant (Feynman-like) gauge in which this approximation becomes reliable \cite{Gusynin563}.

One can show that the solution of  (\ref{auto1}) is such that the magnetic field catalyzes the dynamical generation of a mass and an AMM even at the weakest attractive interaction \cite{ferrerincera}. This is the realization of the MC$\chi$SB \cite{severalaspects} . Carrying out the $\mathbb{E}_p$ transformation in  \Eq{auto1} and taking into account Eqs. (\ref{ad10}) and (\ref{full-FP}) we obtain
\bea
     \hat{\delta}_{ll'} \widetilde{\Sigma}^l(\overline{p})\Pi(l)&=&
      -ie_{R}^2\sum_{l''}
       \int \frac{d^4q}{(2\pi)^4}
       \frac{e^{-\hat{q}^2_\perp}}{q^2}
    \sum_{[\sigma]}
    \frac{
         e^{i(n-n''+\overline{n}''-n')\phi}
        J_{nn''}(\hat{q}_\perp)J_{\overline{n}''n'}(\hat{q}_\perp)
         }
         {\sqrt{n!n''!n'!\overline{n}''!}}
\nonumber \\
    &&\hspace{1.5cm}\times
{
    \Delta(\sigma)\gamma^\mu\Delta(\sigma'')\Pi(l'')
         \tilde{G}^{l''}(\overline{p-q})
    \Delta(\overline{\sigma}'')\gamma_\mu\Delta(\sigma')}
\nonumber \\
\label{auto8}
\eea
with $n=n(l,\sigma)$, $n'=n(l',\sigma')$, $n''=n(l'',\sigma'')$,
$\overline{n}''=n(l'',\overline{\sigma}'')$, $[\sigma]$ meaning sum over $\sigma$, $\sigma^\prime$, $\sigma^{''}$ and $\bar\sigma^{''}$. In the above result we used
\bea
   D^{\mu\nu}(x-y)=\int\frac{d^4q}{(2\pi^4)}
       e^{-iq\cdot (x-y)}\frac{g^{\mu\nu}}{q^2}
\label{auto3}
\eea
 as well as \cite{leeleungng}
\bea
  &&\hspace{-1.5cm}\int dy e^{-i q'\cdot y}
    \overline{\mathbb{E}}_p(y)\gamma_\nu
    \mathbb{E}_{p'}(y)
=   (2\pi)^4 \hat{\delta}^{(3)}(p'+q'-p)
   e^{-\frac{\hat{q}'^2}{2}}
   e^{-i\frac{q_1'(p_2'+p_2)}{2eH}}
    \sum_{\sigma, \sigma'}\frac{J_{nn'}(\hat{q}'_\perp)e^{i(n-n')\phi}}{\sqrt{n!n'!}}
{    \Delta(\sigma)\gamma_\nu\Delta(\sigma')}
\label{oppol6}
\eea
and
\bea
   &&\hspace{-1.5cm}\int dx e^{i q\cdot x}
     \overline{\mathbb{E}}_{p'}(x)\gamma_\mu
     \mathbb{E}_{p}(x)
=    (2\pi)^4 \hat{\delta}^{(3)}(p'+q-p)
    e^{-\frac{\hat{q}^2}{2}}
    e^{i\frac{q_1(p_2'+p_2)}{2eH}}
    \sum_{\overline{\sigma},\overline{\sigma}'}
    \frac{J_{\overline{n}'\overline{n}}(\hat{q}_\perp)e^{i(\overline{n}'-\overline{n})\phi}}
    {\sqrt{\overline{n}!\overline{n}'!}}
{    \Delta(\overline{\sigma}')\gamma_\mu\Delta(\overline{\sigma})},
\label{oppol7}
\eea
where
\bea
    J_{nn'}(\hat{q}_\perp)
     \equiv\sum_{m=0}^{min(n,n')}
           \frac{n!n'!\,\,|i\hat{q}_\perp|^{n+n'-2m}}{m!(n-m)!(n'-m)!}\
\label{jeqn}
\eea
and $\hat{q}_\perp\equiv q_\perp/2eH$ is the normalized transverse momentum.

Taking into account that in a strong magnetic field, the exponential
factor in \Eq{auto8} serves as a cutoff for large transverse momentum, the main contribution
to the fermion self-energy comes from the infrared region, $\hat{q}_\bot\ll1$.
It is then justified the approximation $J_{nn'}(\hat{q}_\perp)\simeq n!\delta_{nn'}$. Hence, the
electron self-energy (\ref{auto8}) in the strong-field approximation
is given by
\bea
     \hat{\delta}_{ll'} \widetilde{\Sigma}^l(\overline{p})\Pi(l)=
      -ie_{R}^2\sum_{l''}\sum_{[\sigma]}
       \int \frac{d^4q}{(2\pi)^4}
       \frac{e^{-\hat{q}^2_\perp}}{q^2}
{
   \Delta(\sigma)\gamma^\mu\Delta(\sigma'')}\Pi(l'')
         \tilde{G}^{l''}(\overline{p-q})
{
    \Delta(\overline{\sigma}'')\gamma_\mu\Delta(\sigma')}
    \delta_{nn''}\delta_{\overline{n}''n'}\,.
\label{auto9}
\eea
Now, using the relation $\delta_{nn''}=\delta_{\sigma\sigma''}\delta_{ll''}
+\delta_{-\sigma\sigma''}\delta_{l''l+\sigma}$ and
performing the sum over $l''$ and $[\sigma]$, we get
\bea
    &&\hspace{-0.2cm} \widetilde{\Sigma}^l(\overline{p})\Pi(l)=
      -ie_{R}^2
    \int \frac{d^4q}{(2\pi)^4}
    e^{-\hat{q}^2_\perp}
    \left\{
    \Pi(l)
    \frac{
    \gamma^\mu_{||}
    \tilde{G}^{l}(\overline{p-q})
    \gamma_\mu^{||}}
    {q^2}
\right.
    +\left.\sum_{\sigma\pm1}
      \frac{\Delta(\sigma)
      \gamma^\mu_{\perp}
      \Pi(l+\sigma)
      \tilde{G}^{l+\sigma}(\overline{p-q})
      \gamma_\mu^{\perp}\Delta(\sigma)}
      {q^2}
      \right\}
\label{auto17}
\eea
Because the self-energy for a given Landau level $l$ in \Eq{auto17}
receives contributions from  the fermion propagators depending on the
Landau level $l$, together with the adjacent ones $l-1$ and $l+1$, the SDE's
for all LLs form an infinite system of couple
equations. However, the problem can be simplified taking into account
that at strong magnetic fields, the leading
contribution to each equation comes from the propagators with the
lowest LLs, since the term $\sim lH$ acts as a suppression factor in
the denominators of the fermion propagators with $l\geq 1$. Under this
approximation, one can find a consistent solution for each
level \cite{ferrerincera}. Moreover, since all the solutions can be ultimately related to
the LLL, which gives the main contribution, we have that all the
dynamical quantities are determined by the LLL infrared dynamics.

The above considerations imply that the SDE for the LLL reduces to
\bea
    \widetilde{\Sigma}^0(\overline{p})\Delta(+)&\simeq&
      -ie_{R}^2
    \Delta(+)
    \int \frac{d^4q}{(2\pi)^4}
       \frac{e^{-\hat{q}^2_\perp}}{q^2}
    \gamma^\mu_{||}
    \tilde{G}^{0}(\overline{p-q})
    \gamma_\mu^{||}.
\label{sd2}
\eea
Substituting the LHS of (\ref{sd2}) with \Eq{ad11}, we obtain that
$Z_{||}^0=0$ and
\bea
   {\mathcal{E}^0}\simeq-i2e_{R}^2
     \int \frac{d^4q}{(2\pi)^4}
      \frac{e^{-\hat{q}^2_\perp}}{q^2}
      \frac{{\mathcal{E}^0}}{(p-q)_{||}^2-{\mathcal{E}^0}^2},
\label{sd9c}
\eea
where we have used that ${\gamma^{||}}^\mu\Lambda^{\pm}_{||}(p)\gamma_\mu^{||}=1$ and introduced the notation for the rest energy $\mathcal{E}^0=M^0+T^0$.  Equation (\ref{sd9c}) coincides with the one found in \cite{NPB462} for the dynamical mass $m_{dyn}$, because at the LLL, due to the lack of spin degeneracy, there is no way to separately determine the dynamical parameters $M^0$ and $T^0$,  as they only enter in the LLL equation through the combination $\mathcal{E}^0=M^0+T^0$. One can show \cite{NPB462} that the dynamical parameter $\mathcal{E}^0$, is independent of the momentum in the infrared region $|p^2| \ll {\mathcal{E}^0}^2 \ll |eH|$, and rapidly decreases, ${\mathcal{E}^0}^2\sim \frac{1}{p^2}$, as $p^2 \to \infty $. This kind of behavior is characteristic of many dynamically induced solutions of SDEs.  For example, it was also found for the dynamical mass in a theory with fermions coupled to a scalar field through a Yukawa vertex \cite{Elizalde}.  

The momentum dependence of the dynamical parameter $\mathcal{E}^0$ indicates that the main contribution to the solution comes from the infrared region. This allows us to set $p_{||}=0$, cut the integral at the scale $\sqrt{eH}$, and assume that $\mathcal{E}^0$ is momentum independent.  After Wick rotating to Euclidean space, the LLL SDE becomes
\bea
   1&\simeq&\frac{\alpha_R}{2\pi}
    \int \frac{d^4q}{(2\pi)^2}\frac{e^{-\hat{q}^2_\perp}}{q^2}
         \frac{1}{ q_{||}^2+{\mathcal{E}^0}^2 }\,,
\label{sd10}
\eea
whose solution is given by
\bea \label{sd10b}
   {\mathcal{E}^0}\simeq\sqrt{2eH}
            e^{-\sqrt{\frac{\pi}{\alpha_{R}}}}
\eea
where $\alpha_R$ is the renormalized fine structure constant at the scale $\sqrt{eH}$. To keep our notation simple we did not write the subindex "R" in the electric charge  $"e"$ in the above expression, but from now on,  in all the formulas related to the nonperturbative theory, $"e"$  should be understood as the renormalized charge at the scale $\sqrt{eH}$. The non-perturbative
character of the solution (\ref{sd10b}) is reflected in its dependence on $\alpha_R$. For higher LL's, each parameter, the dynamical mass and the AMM, can
be found as functions of $\mathcal{E}^0$ \cite{ferrerincera}. They  are all smaller than $\mathcal{E}^0$, reflecting the fact that the MC$\chi$SB is an infrared phenomenon essentially determined by the LLL contribution.  We will use the solution (\ref{sd10b}) to show that the chirally broken phase of massless QED in a magnetic field behaves as a paraelectric medium.

\section{Non-Perturbative Euler-Heisenberg Lagrangian for Massless QED}~\label{EulerHeisenberSec}

To investigate the electric response of massless QED in a strong magnetic field we are going to
find the corresponding non-perturbative Euler-Heisenberg Lagrangian. The non-perturbative character of this effective Lagrangian comes from the fact that we are going to use the ladder full electron propagator considered in the previous section. Therefore, the mass and AMM parameters in our formulas should be understood as the dynamical solutions of the ladder SDEs. In all the derivations that follow, the ground state of the system is driven by the strong magnetic field, while the electric field just plays the role of a weak probe.

The non-perturbative Euler-Heisenberg action $\Phi_{eff}[A]$  can be found from the following path integral 
\bea
  e^{i\Phi_{eff}[A]}=
  \int [\mathcal{D}\{\overline{\psi}\}][\mathcal{D}\{\psi\}]
       e^{iS[\{\overline{\psi}\},\{\psi\},A_\mu]}
\label{effecactionA}
\eea
where
\bea
   S[\{\overline{\psi}\},\{\psi\},A_\mu]&=&
   \int d^4x
   \left[
   \overline{\psi}(x)
   \left[\gamma^\mu\Pi_\mu-\Sigma(x)\right]
   \psi(x)
   -\frac{1}{4}F^{\mu\nu}F_{\mu\nu}
   \right]
\label{QEDaction}
\eea
is the QED action in the presence of an external electromagnetic potential $A_\mu$. Notice that it includes the
electron self-energy $\Sigma(x)$, which in our case is the solution of Eq.~(\ref{auto1}) and thus a function of the dynamical parameters characterizing the phase with chiral symmetry breaking.
Therefore, the effective action contains the correction to the classical QED action that comes from the resummation of the infinite number of rainbow diagrams entering in the ladder approximation for  $\Sigma(x)$. To obtain an explicit expression for $\Phi_{eff}$, we take (\ref{QEDaction}) as the initial action of an effective  theory on which the gauge field fluctuations are neglected, and hence the electromagnetic field is merely reduced to a background field.  

In order to simplify the
calculations, it is common to expand the effective action in powers of
$\hbar$. In the presence of an external electromagnetic field,
the effective action can be expanded up to linear terms in $\hbar$, where,  since we are dealing with a non-perturbative theory, all the internal fermion lines are taken as the full fermion propagator. Then,
\bea
   \Phi_{eff}=\Phi^{(0)}+\Phi^{(1)}
\label{contrib}
\eea
with
\bea
   \Phi^{(0)}=-\frac{1}{4}\int d^4x\ F^{\mu\nu}F_{\mu\nu}
\label{0contrib}
\eea
and
\bea
  e^{i\Phi^{1}[A]}=
  \int [\mathcal{D}\{\overline{\psi}\}][\mathcal{D}\{\psi\}]
       e^{i\int d^4x\ \overline{\psi}(x)[\gamma^\mu\Pi_\mu-\Sigma(x)]\psi(x)}
\label{defW1}
\eea
The Feynman diagram associated with (\ref{defW1}) consists in principle of a fermion
bubble with  infinite insertions of photon fields~\cite{Dunne}, which in this approach corresponds to external field lines. To investigate the linear electric response of the medium in a strong magnetic field, we are going to consider that in addition to the strong magnetic field, there is a weak but nonzero electric field. This last one can serve as a probe to explore the electric susceptibility of the system described by the action (\ref{defW1}). In Ref. \cite{prlFIS}, we proved that the confinement of the electrons to the LLL by the strong magnetic field leads  to an anisotropic electric  susceptibility. In the plane transverse to the magnetic field direction the susceptibility was just the same as in vacuum. Here we are interested in exploring the electric susceptibility in the direction along the magnetic field.  With that aim we can consider constant parallel electric and magnetic fields in the $x_3$-direction. In the Coulomb gauge they are described by the photon field $A_\mu=(-Ex_3, 0, Hx_1, 0)$.

Performing a Wick rotation ($x_0\rightarrow -i x_4$, $\gamma_0 \rightarrow -i \gamma_4$, and $E \rightarrow iE$) in \Eq{defW1} to go to
Euclidean  variables we obtain
\bea
  e^{\Phi^{1}_E[A]}=
  \int [\mathcal{D}\{\overline{\psi}\}][\mathcal{D}\{\psi\}]
       e^{\int d^4x_E\ \overline{\psi}(x)[-\gamma_\mu\Pi_\mu-\Sigma(x)]\psi(x)}
\label{defW1eucl}
\eea
where for the sake of simplicity, we keep the same notation for the $\gamma$ matrices, etc,  but it is understood from now on that they are all Wick-rotated.

In order to transform (\ref{defW1eucl}) to momentum space, it is convenient to use Ritus's approach and transform the spinors with the eigenfunctions $\mathbb{E}_p$ of the parallel fields case,
\bea
   \psi(x)=\sumint \frac{d^4p}{(2\pi)^4}\mathbb{E}_p(x)\psi(p)\ ,  \ \ \
   \overline{\psi}(x)=\sumint\frac{d^4p}{(2\pi)^4}
                      \overline{\psi}(p)\overline{\mathbb{E}}_p(x).
\label{spinortransform}
\eea

Details of the calculation of the $\mathbb{E}_p$ functions for constant and parallel electric and magnetic fields in Euclidean space can be found in Appendix~\ref{ritusappendix}.  There,  we also verify some important properties satisfied by these functions.

Next, using Eqs.~(\ref{spinortransform}), (\ref{epsep}),
(\ref{orthonormalityfull}) and~(\ref{mainproperty}), we can rewrite the Euclidean QED action
(\ref{defW1eucl}) in momentum space as

\bea
   \int d^4x_E\ \overline{\psi}(x)[-\gamma_\mu\Pi_\mu-\Sigma(x)]\psi(x)&=&
   \int d^4x_E\sumint \frac{d^4pd^4p'}{(2\pi)^8}
   \overline{\psi}(p)\overline{\mathbb{E}}_{p}(x)
   [-\gamma_\mu\Pi_\mu-\Sigma(x)]
   \mathbb{E}_{p'}(x)\psi(p')
\nonumber \\
   &=&
   \sumint \frac{d^4p}{(2\pi)^4}
   \overline{\psi}(p)
   \Pi^{E}(\tilde{l})\Pi^H(l)
   [-\gamma_\mu\overline{p}_\mu-\widetilde{\Sigma}(\overline{p})]
   \psi(p)
\label{actioneuclritus}
\eea
Here we took advantage of the fact that the self-energy $\widetilde{\Sigma}(\overline{p})$ is
diagonal in the basis spanned by the $\mathbb{E}_p$ functions.  We assume that the self-energy has the same structure as in the case with a pure magnetic field. This is a reasonable assumption, as we are interested in the situation of a strong magnetic field, but a very weak electric field, so the ground state of the system should not be affected by the presence of the electric field. Accordingly, the self-energy structure is still given by Eq. (\ref{ad11}), but with the components of the momentum $\overline{p}$ defined by
  $\overline{p}_\mu^\|=(sgn(eE)\sqrt{2|eE|\tilde{l}},0,0,0)$ and
  $\overline{p}_\mu^\bot=(0,0,sgn(eH)\sqrt{2|eH|l},0)$.

Carrying out the functional integrals  in
\Eq{defW1eucl}, the Euclidean 1-loop effective
action turns out to be of the form
\bea
\phi^{1}_E[A]
  = \frac{|eH||eE|}{2(2\pi)^2}\sum_{\overline{l},l=0}^\infty
   \sum_{\sigma=\pm1}
   \ln\left[V^2(\overline{p})
  +(M^{l}+\sigma T^{l})^2\right]
       \left(1-\frac{1-\hat{\sigma}}{2}\delta_{l0}\right)
       \left(2-\delta_{\tilde{l}0}\right)
\label{W1}
\eea
with $V_\mu=(1+Z^{l}_{\|})(sgn(eE)\sqrt{2|eE|\tilde{l}},0,(1+Z^{l}_{\perp})sgn(eH)\sqrt{2|eH|l},0)$,  $\phi^1_E[A]\equiv(L_4)^{-1}\Phi^1_E[A]$, and  $L_4$ denoting the
four-dimensional volume in Euclidean space. We call attention to the convenience of using Ritus's formalism to
incorporate the non-perturbative effects coming from the
self-energy $\Sigma(\overline{p})$ in momentum space in a straightforward and simple way.

As discussed in Sec. II, the ground state of the magnetically catalyzed system is mainly determined by the infrared dynamics of the fermions lying in the LLL. Out of this region, the LLL solution $\mathcal{E}^0$ and with it,  the rest of the dynamical parameters, quickly go to zero.  Hence, it is consistent to limit the sum in $l$ to the region where the chiral symmetry is dynamically broken, take the dynamical parameters as momentum-independent in that region, and just keep the leading contribution given by the LLL term $(l=0, \sigma= +)$ in (\ref{W1}). Using the Schwinger's proper time representation for the logarithm and summing in $\overline{l}$, we find
\bea
    (\phi^{1}_E)[A]&=&
       \frac{1}{8\pi^2}
       \int_{1/|eH|}^\infty\frac{ds}{s}
       e^{-s{(\mathcal{E}^0})^2}
       |eH||eE|
       \coth(|eE|s)
\label{w1HstrongEweak}
\eea
where $\mathcal{E}^0$  is given by \Eq{sd10b}. Notice that the consistency of the LLL approximation requires to introduce an ultraviolet cutoff $1/|eH|$ in the $s$ integration. This is in agreement with the fact, previously stressed, that the nonperturbative MC$\chi$SM is essentially an infrared phenomenon that takes place for energies below the scale $\sqrt{eH}$. For energies larger than this scale the dynamically induced rest energy goes to zero \cite{NPB462}, and the expression (\ref{w1HstrongEweak}) is not valid.

Let us make contact here with the usual Euler-Heisenberg Lagrangian
of perturbative massive QED in external electromagnetic fields. It
can be obtained from (\ref{W1}) by replacing the dynamical
parameters by the electron mass (i.e. $M^{l}+\sigma
T^{l} \rightarrow m$), introducing the Schwinger's proper time, and subtracting the vacuum contribution to
regularize the divergence at zero fields,
\bea
  (\phi^1_E)^{reg}_{QED}[A]
    &=&
       \frac{1}{8\pi^2}
       \int_0^\infty\frac{ds}{s^3}
       e^{-s m^2}
       \left[
       |eEs||eHs|
       \coth(|eE|s)\coth(|eH|s)
      -1
       \right]
\label{EulerHeisenbergren}
\eea

The effective action (\ref{EulerHeisenbergren}) still has a field-dependent ultraviolet logarithmic divergence, which becomes
evident in the expansion of the hyperbolic cotangents in the squared parenthesis. Following Ref.~\cite{Schwinger},  the
renormalization procedure is achieved by renormalizing the charge and fields and subtracting in the effective action the counter term,
\bea
  (\phi^1_E)_{ct}^{ren}[A]
    &=&
       \frac{1}{8\pi^2}
       \int_0^\infty\frac{ds}{s^3}
       e^{-s m^2}
       \frac{(|eE|s)^2+(|eH|s)^2}{3},
\label{actionren}
\eea
where the charge and fields appearing in Eqs. (\ref{EulerHeisenbergren})-(\ref{actionren}) should be understood as the renormalized
ones.

Subtracting (\ref{actionren}) from (\ref{EulerHeisenbergren}), and doing the analytical continuation  $E\rightarrow -iE$, we recover the well-known
Euler-Heisenberg Lagrangian of massive QED in Minkowski
space~\cite{QED-lifshitz}. 

The path integral formulation of the effective action for massive and
massless QED in the presence of a constant electromagnetic field has
been previously found in Refs.~\cite{soldati,bassetto} in terms of the
spectra of the Euclidean Dirac's operator and the generalized
$\zeta$-function, but within a perturbative approach. A perturbative
approach does not allow to incorporate the phenomenon of MC$\chi$SB.
However, the MC$\chi$SB is unavoidable in the massless case, because
once a magnetic field is present,  no critical magnetic
strength is required for the dynamical breaking of the chiral symmetry to
occur and for the mass and AMM to be generated.

\section{Paraelectricity}~\label{paraelecsec}
With the help of Eq. (\ref{w1HstrongEweak}) we can now show the paramagnetic behavior of massless QED under an applied magnetic field. First, notice that the electric polarization of the system can be obtained as minus the derivative of the effective action, or
electromagnetic free-energy density $\Phi$ with respect to an applied electric
field. For a weak electric field E, the free-energy density can be
expanded in powers of E as
\bea
  \Phi=\Phi_0-\eta E-\frac{1}{2}\chi E^2.
\label{freeenergy}
\eea
In a strongly magnetized medium, the coefficients $\Phi_0$, $\eta$, $\chi$, etc.,
may in principle depend on the magnetic field. The susceptibility
coefficient $\eta$ is different from zero for ferroelectric
materials \cite{ferroelectricity}. In magnetized QED it is zero,
because the second term in the RHS of \Eq{freeenergy}
violates parity, a symmetry that is not broken neither in
massive QED nor in the chirally broken phase of massless
QED. Then, the coefficient $\chi$ characterizes the lowest order of
the system dielectric response. It accounts for the electric
polarization $P=\chi_{E} E=\frac{\partial \Phi}{\partial E}$ induced by an externally applied electric field.
This term does not break any additional symmetry of the free energy

Assuming that $E \ll {(\mathcal{E}^0})^2$,
we can perform a Taylor expansion of $(|eE|s) \coth(|eE|s)$ in power of
$|eE|/{(\mathcal{E}^0})^2$ in (\ref{w1HstrongEweak}).  Keeping just the leading term in the expansion, the effective
action can be approximated by
\bea
  (\phi^{1}_E)[A]
     &\approx&
       \frac{|eH|}{8\pi^2}
       \int_{1/eH}^\infty \frac{ds}{s^2}
       e^{-s{(\mathcal{E}^0)}^2}
       \left[\frac{(|eE|s)^2}{3}\right]
        \approx
       \frac{|eH|(|eE|)^2}{24\pi^2}
       \int_{1/eH}^{1/{\mathcal{E}^0}^{2}}ds
\nonumber \\
      &=&
      \frac{\alpha_{R}}{6\pi}\frac{|eH|}{({\mathcal{E}^0})^2} E^2
\label{w1HstrongEweak2}
\eea
Note that the Feynman diagram associated with this equation
corresponds to the photon polarization operator with internal fermion lines of full fermion propagators,  two external legs
of ``electric'' photons, and an infinite number of insertions of
``magnetic'' photons. A straightforward calculation of the
polarization-operator $\Pi^{00}$ in the static limit corroborates the
above observation~\cite{Shabad-1,coefficient-LLL,prlFIS}.

To regain the effective action in Minkowski space we should make the analytic continuation, $E \rightarrow - iE$, in (\ref{w1HstrongEweak2}). Comparing the obtained action with Eq.(\ref{freeenergy}), we immediately find that for the magnetically catalyzed QED the electric susceptibility in the direction of the magnetic field is
\bea
  \chi_E=
      \frac{\alpha_{R}}{3\pi}\frac{|eH|}{({\mathcal{E}^0})^2}
\label{w1HstrongEweak2-1}
\eea
Using (\ref{sd10b})  in (\ref{w1HstrongEweak2-1}) we obtain
\bea
   \chi_E=\frac{\alpha_{R}}{6\pi}e^{\sqrt{\frac{\pi}{\alpha_{R}}}},
\label{susceptibility}
\eea
in agreement with the result found in the infrared limit from the photon polarization operator in Ref. \cite{prlFIS}. The non-perturbative dependence of the susceptibility on the renormalized fine-structure constant $\alpha_{R}$ accounts for the large electric response along the field direction of the medium with MC$\chi$SB. The susceptibility in the transverse direction was found in \cite{prlFIS}. It resulted to be zero as in vacuum. Therefore, at strong magnetic field the system with MC$\chi$SB displays a noticeable anisotropy in its electric response.

It is worth to point out here the difference between the electric responses of the MC$\chi$SB system and of the regular massive QED in a strong magnetic field. In the massive QED case, the  Euler-Heisenberg action is found perturbatively. The one-loop diagrams contributing to the perturbative action contain a logarithmic divergence that needs to be regularized, as pointed out in the previous section. This ultraviolet divergence affects the charge  and field renormalization and leads to a logarithmic contribution $\sim \alpha (E^{2}-H^{2})\ln(eH/m^2)$ to the action, which in turn affects the susceptibility in the parallel and transverse directions  \cite{Ragazzon}. In this situation, the strong field limit can be taken only after the divergences of the theory are eliminated. In the nonperturbative case studied in the present work,  the ultraviolet logarithmic divergence is absent because the dynamics of the fermions that leads to the magnetically catalyzed, chirally broken phase is essentially infrared ($p^2<<eH$),  so the nonperturbative Euler-Heisenberg action has a dynamical cutoff at the scale $\sqrt{eH}$ that limits it to the infrared region where the dynamical parameters are momentum-independent and nonzero. Besides, in the ladder approximation used to obtain the physical parameters, the full photon propagator and vertex reduce to the free ones, so any potentially dangerous infrared divergence is also cancelled out and the electromagnetic field and coupling constant are automatically renormalized at the scale  $\sqrt{eH}$. Hence, in the nonperturbative case the transverse susceptibility remains zero as in the vacuum.
 
It should come as no surprise the different electric behaviors in regular massive QED and in massless QED with MC$\chi$SB. In massive QED, the ground state is driven by unpaired electrons, while in the nonperturbative case, the electrons pair with positrons forming tiny electric dipoles. As the electrons forming the dipoles are all in the LLL, their dynamics is (1+1)- dimensional (their dispersion only depends on the parallel momentum)  due to the well-known dimensional reduction that occurs at the LLL. Hence, the electric dipoles  can be polarized by any small electric field to produce a strong polarization only in the parallel direction. The role of the magnetic field here is to induce the dipole moments, while the role of the electric field is to polarize them.

The lack of explicit magnetic field dependence in the susceptibility (\ref{susceptibility}) is a unique feature of the paraelectricity in massless QED with MC$\chi$SB. This property is not found in other strongly magnetized systems like massive QED, nuclear matter, color superconductivity, etc. \cite{Bo}. Even in the case of MC$\chi$SB in the context of QCD in the presence of a strong magnetic field, the
\textit{chromo}-susceptibility remains an explicit function of the magnetic
field through the running of the strong coupling $\alpha_s$
\cite{Igor}.

\section{Conclusion}~\label{conclusion}

In this paper we obtained the non-perturbative Euler-Heisenberg Lagrangian for massless QED with MC$\chi$SB. The magnetic field catalyzes the breaking of chiral symmetry through the generation of the dynamical parameters that are solutions of the non-perturbative Schwinger-Dyson equations. Since the dynamics of the electron-positron pairing leading to the breaking of the chiral symmetry is mainly induced in the region of momenta much smaller than $\sqrt{eH}$, the magnetic field introduces a dynamical ultraviolet cutoff in the theory that also enters in the non-perturbative Euler-Heisenberg action and provides the scale at which the coupling is renormalized in the calculations.  The electric response of the medium with MC$\chi$SB,  characterized by the electric susceptibility,  can be obtained from the Euler-Heisenberg Lagrangian.
The non-perturbative character of the calculation is reflected in the dependence of the susceptibility on the coupling constant. A remarkably feature of the electric response is its anisotropy. In the direction parallel to the magnetic field, the susceptibility is large and independent of the magnetic-field strength, while in the direction transverse to the field it is zero. Given that the amplitude of the dynamical parameters are typically too small to be observable in the experiment, the large electric susceptibility is the best candidate available to test
the realization of the $MC\chi SB$ phenomenon in a given physical system.

An important implication of this result is that the chirally broken
phase exhibits strong paraelectricity, a property found in certain
condensed matter systems like quantum paraelectric (QP) materials
~\cite{Paraelectricity} and transition-metal-oxides (TMO)
\cite{TMO}. In those materials, unaligned electric dipoles are aligned
in an external electric field, producing a high electric
susceptibility, often exceeding $10^4$. In QP materials the large
electric susceptibility is temperature-independent below certain
critical temperature, a property attributed to a quantum phase
transition~\cite{Paraelectricity}. An interesting question to explore
in the future is whether the strong susceptibility found here within a
(3+1)-dimensional theory is also present in quasiplanar condensed
matter systems as bilayer graphene. It is known, that the band
structure of bilayer graphene can be controlled by an applied electric
field perpendicular to the layers' plane. The electric field creates
an electronic gap between the valence and conduction bands with energy
values that varies from zero to mid-infrared~\cite{Castro}, depending
on the field strength. Under a very weak electric field the gap is
practically zero and the spectrum is Dirac-like. Even though this is a
very peculiar (3+1)-D system, only formed by two layers, one could
attempt to model it with a (3+1)-D theory of interactive massless
fermions. Due to the universality of the MC$\chi$SB, we expect that
the application of a strong magnetic field parallel to the weak
electric one will trigger the generation of a dynamical energy
gap. Under these conditions, one would expect that detecting a very
large electric susceptibility in the direction of the applied fields
would signal the realization of the MC$\chi$SB phenomenon.

Another interesting direction worth to be explored in the future within the framework of the non-perturbative Euler-Heisenberg Lagrangian is in connection with higher order non-linear effects. Recently \cite{Loewe}, the study of higher order nonlinear effects in the context of the perturbative QED Euler-Heisenberg Lagrangian indicated that a purely magnetic moment placed in an external quasistatic electric field can lead to the induction of an electric moment and viceversa. It would be interesting to investigate if a similar nonlinear effect occurs in the case of the 
magnetically catalyzed system.  

\acknowledgments

We thank Bo Feng for useful discussions and A. E. Shabad for suggesting us to find the susceptibility of the non-perturbative system through the Euler-Heisenberg approach and for enlightening discussions. This work has been supported in part by DOE Nuclear Theory grant DE-SC0002179.

\appendix

\section{Ritus $\mathbb{E}_p$ functions for parallel electric and
  magnetic fields in Euclidean space} \label{ritusappendix}
\subsection{Euclidean $\mathbb{E}_p$ functions}
In this Appendix, we find the matrix eigenfunctions $\mathbb{E}_p$ for
charged fermion in the presence of constant electric and
magnetic fields. They can be used to diagonalize the full fermion
propagator in momentum space. This approach has been extensively used
in the literature for the case of a constant magnetic field, where the
$\mathbb{E}_p$ functions play the role, in a magnetic field,  of the
plane waves (Fourier transform) at zero field. This is the essence of
the so-called Ritus' method \cite{Ritus:1978cj}. Here, we
will consider this approach for the case of constant and parallel
electric and magnetic fields, $\overrightarrow{E}\cdot
\overrightarrow{B}\neq0$, $\overrightarrow{E}\times
\overrightarrow{B}=0$ in Euclidean space, where these
fields enter in a very symmetric way.

In the presence of electric and magnetic fields, the self-energy
operator is a combination of four scalar operators:
$\gamma_\mu\Pi^\mu$, $\sigma_{\mu\nu}F^{\mu\nu}$,
$(F^{\mu\nu}\Pi_\mu)^2$, and $\gamma_5\tilde{F}^{\mu\nu}F_{\mu\nu}$,
with $F_{\mu\nu}=\partial_{\mu}A_{\nu}-\partial_{\nu}A_{\mu}$,
$\tilde{F}^{\mu\nu}=\frac{1}{2}\epsilon_{\mu\nu\alpha\tau}F^{\alpha\tau}$,
$\sigma_{\mu\nu}=i[\gamma_\mu,\gamma_\nu]/2$, and
$\Pi_\mu=(i\partial_\mu-eA_\mu)$. Since all these scalars commute with
$(\gamma_\mu \Pi^\mu)^2$,  the self-energy, and hence the fermion
propagator, will be diagonal in the basis spanned by the
eigenfunctions of  $(\gamma_\mu \Pi^\mu)^2$.

Without loss of generality, we assume that the fields
point along the positive $x_3$-axis.  We fix the gauge,  $A_\mu=(-E
x_3,0,H x_1,0)$, for the external field, and work in the metric
$g_{\mu\nu}=(1,-\overrightarrow{1})$. Taking into account that
$[\Pi_\mu,\Pi_\nu]=-ieF_{\mu\nu}$, one can write the operator
$(\gamma_\mu \Pi^\mu)^2$ as

\bea
   \slsh{\Pi}^2&=&
  \Pi^2-\frac{e}{2}\sigma^{\mu\nu}F_{\mu\nu}
   =\Pi_0^2-\Pi_1^2-\Pi_2^2-\Pi_3^2
           -e\left(\sigma^{03}F_{03}+\sigma^{12}F_{12}\right)
\nonumber \\
&=&
   (i\partial_0+eEx_3)^2-(i\partial_1)^2-(i\partial_2-eHx_1)^2
    -(i\partial_3)^2-e\left(iE\gamma_5\Sigma_3-\Sigma_{3}H\right)
\label{pisquareminkowski}
\eea

Working in the chiral representation of the gamma matrices
\bea
   \gamma_0=\left(
     \begin{array}{cc}
      0 & -1 \\
      -1 & 0
     \end{array}\right),
\hspace{1cm}
   \gamma_i=\left(
     \begin{array}{cc}
      0 & \sigma^i \\
      -\sigma^i & 0
     \end{array}\right),
\hspace{1cm}\mbox{and}\hspace{1cm}
   \gamma_5=\left(
     \begin{array}{cc}
      1 & 0 \\
      0 & -1
     \end{array}\right)
\label{diracgammaMink}
\eea
on which both  $\Sigma_3 =i\gamma_{1}\gamma_{2} $  and  $\gamma_5\equiv i\gamma_0\gamma_1\gamma_2\gamma_3$ are diagonal, the eigenfunctions of the square Dirac operator eigenvalue problem
\bea
 (\gamma_\mu \Pi^\mu)^2\varphi=p^{2}\varphi
 \label{KGequMink}
\eea
 can be written as
\bea
\varphi=E_{p,\sigma, r}(x)\nu_{\sigma, r},
\label{eigenfunctionform}
\eea
where the functional part $E_{p,\sigma, r}(x)$ of the eigenfunction has to be found from the eigenvalue equation (\ref{KGequMink}),  while the spinors  $\nu_{\sigma r}$ are chosen as the eigenvectors of the spin $\Sigma_{3}$ and electric dipole $\gamma_{5}\Sigma_{3}$ operators, i.e., $\Sigma_{3}\nu_{\sigma,r}=\sigma \nu_{\sigma,r}$, $\gamma_{5}\Sigma_{3}\nu_{\sigma,r}=r \nu_{\sigma,r}$, with $\sigma=\pm1$ and $r=\pm1$.

Since we will use the $\mathbb{E}_p$ functions in the
framework of the path-integral formulation, we just need to find them in Euclidean space. With this aim, we perform a Wick
rotation $x_0 \rightarrow -i x_4$,  so that the Euclidean eigenvalue equation becomes
\bea
    \left[(-i\partial_4+eEx_3)^2
    +(-i\partial_1)^2+(-i\partial_2-eHx_1)^2
    +(-i\partial_3)^2-e\left(rE+\sigma H\right)
      \right]E_{p,\sigma,r}(x)
    =p^{2} E_{p,\sigma,r}(x)
\label{euclideanver}
\eea
To solve \Eq{euclideanver} we separate in longitudinal and
perpendicular variables via
\bea
 E_{p,\sigma,r}(x)=E^{E}_{p,r}(x_{\|})E^{H}_{p,\sigma}(x_{\perp})
\label{epsep}
\eea
where
\bea
E^{E}_{p,r}(x_{\|})\equiv
 e^{ip_4x_4}
 \chi_{p,r}^{E}(x_3)
 \label{epelectric}
 \eea
 \bea
E^{H}_{p,\sigma}(x_{\perp})
  &\equiv&
  e^{ip_2x_2}
 \chi_{p,\sigma}^{H}(x_1)
\label{epmagnetic}
\eea
and substitute \Eq{epsep} into  \Eq{euclideanver} to obtain an eigenvalue equation for each $\chi$ function. The one for $ \chi_{p,\sigma}^{H}(x_1)$ is
\bea
   \left[-\partial_1^2+(p_2-eHx_1)^2-(p_{\perp}^2+eH\sigma\right)]\chi_{p,\sigma}^H(x_1)=0
\label{magneticequation}
\eea
where the eigenvalues $p_{\perp}^2\equiv p^2 -p_{\|}^2$, and $p_{\|}^2$ are the constants usually introduced in the separation of variable method that has to be determined from the eigenvalue equations.  \Eq{magneticequation} is the same as the one found in the case of a system with just a magnetic field  \cite{leeleungng}. Its solutions are the parabolic cylinder functions $D_{n_{H}}(\rho_H)$ with argument $\rho_{H}=\sqrt{2|eH|}\left(\frac{p_2}{eH}-x_1\right)$ and index
\bea
n_{H}=n_{H}(\sigma,l)\equiv l+\emph{sgn}(eH)\frac{\sigma}{2}-\frac{1}{2},    \quad \quad n_{H}=0,1,2,...
\label{nH}
\eea
The non-negative integer $l$ is the Landau level,  which as known characterizes the quantization of the transverse momentum in a magnetic field. Notice that in the LLL,  $l=0$, only one spin projection is allowed, that is, $ \sigma=+1$ if $\emph{sgn}(eH)>0$, or $ \sigma=-1$ if $\emph{sgn}(eH)<0$. The eigenvalue  satisfies $p_{\perp}^2=p^2 -p_{\|}^2 =2|eH|l$.

The equation for $ \chi_{p,r}^{E}(x_1)$ is
\bea
     \left[-\partial_3^2+(p_4+eEx_3)^2
    -({p}_{||}^2+eEr)
      \right]\chi_{p,r}^{E}(x_3)
    =0
 \label{e-quation}
\eea
which, as \Eq{magneticequation}, is also the parabolic cylinder equation and hence has solutions of the form $\chi_{p,r}^{E}(x_3)=D_{n_{E}}(\rho_E)$ with argument $\rho_{E}=\sqrt{2|eE|}\left(x_3+\frac{p_4}{eE}\right)$, and index
\bea
n_{E}=n_{E}(r,\widetilde{l})\equiv \widetilde{l}+\emph{sgn}(eE)\frac{r}{2}-\frac{1}{2},    \quad \quad n_{E}=0,1,2,...
\label{nE}
\eea
On the other hand, the non-negative integer  $\widetilde{l}$ enters in a similar way to the Landau level, but it corresponds to the quantization of the longitudinal momentum in the presence of an electric field in Euclidean space. Therefore, in Euclidean space, there is a symmetry between the electric and magnetic field sectors of the solution. Notice that $r$ represents the projection in the electric field direction of an "intrinsic" electric dipole moment. Only one projection of $r$ is allowed at $\widetilde{l}=0$.

From \Eq{e-quation} one finds $p_{\|}^2 = 2|eE|\widetilde{l}$, so solving for $p^2$ we obtain for the eigenvalue in (\ref{euclideanver}),  $p^{2}=p_{\perp}^{2}+p_{\|}^{2}=2|eH|l+2|eE|\widetilde{l}$. The corresponding normalized eigenfunctions are
\bea
E_{p,\sigma, r}(x)=N_{n_{H}} e^{ip_2x_2} D_{n_{H}}(\rho_H)N_{n_{E}} e^{ip_4x_4} D_{n_{E}}(\rho_E)
\label{finaleigenf}
\eea
with normalization constant $N_{n_H}=\left(4\pi |eH|\right)^{\frac{1}{4}}/\sqrt{n_{H}!}$, and $N_{n_E}$ found by replacing $H$ by $E$ in $N_{n_H}$.

Introducing the spin and electric dipole projectors
\bea
  \Delta^{H}(\sigma)=\frac{1}{2} \left(1+\sigma \Sigma^3\right), \ \ \ \ \
   \Delta^{E}(r)=\frac{1}{2}\left(1+r\gamma_5\Sigma^3 \right)
\label{projectors}
\eea

which satisfy
\bea
  \Delta^{H}(+) \Delta^{H}(-)=0,  \quad \Delta^{E}(+) \Delta^{E}(-)=0
\eea
\bea
\Delta^{H}(+)+ \Delta^{H}(-)=1,  \quad  \Delta^{E}(+) + \Delta^{E}(-)=1
\eea
 \bea
[\Delta^{H}(\sigma),\Delta^{E}(r)]=0
\label{deltalgebra}
\eea
The Euclidean $\mathbb{E}_p$ functions of the problem with parallel constant electric and magnetic fields can be defined as
\bea
   \mathbb{E}_p=\mathbb{E}_p^{E}(x_{||})\mathbb{E}_p^H(x_{\perp})
\label{epfinal}
\eea
with
\bea
\mathbb{E}_p^{E}(x_{||})
 \equiv {\sum_{r=\pm1}}E_{p,r}^{E}(x_\|)
{
  \Delta^{E}(r)}=  e^{ip_4x_4}
  \sum_{r=\pm1}\chi_{p,r}^{E}(x_3)
{
  \Delta^{E}(r)}
\label{epelectricfull}
\eea
\bea
\mathbb{E}_p^{H}(x_{\perp})
 \equiv {\sum_{\sigma=\pm1}}E_{p,\sigma}^{H}(x_\perp)
{
  \Delta^{H}(\sigma)}=
  e^{ip_2x_2}
  {\sum_{\sigma=\pm1}}\chi_{p,\sigma}^{H}(x_1)
{
  \Delta^H(\sigma)}
\label{epmagneticfull}
\eea

It is known that spinors in Euclidean space
  obey
\bea
  \left\{\varphi(x),\varphi(y)\right\} =
  \left\{\overline{\varphi}(x),\overline{\varphi}(y)\right\} =
  \left\{\varphi(x),\overline{\varphi}(y)\right\}=0
\label{coleman56}
\eea
where the last relation implies that $\overline{\varphi}$ is not
necessarily obtained as the product $\varphi^\dagger\gamma_4$. Then, $\varphi$ and $\overline{\varphi}$ result in two
totally independent functions that should be found from two corresponding independent Euclidean equations. This is one of the main
novelty of Euclidean fermion field theory~\cite{coleman}.

Hence, to obtain the Euclidean $\overline{\mathbb{E}}_p$  function, since it is associated with the spinor
$\overline{\varphi}\equiv\overline{\nu}_{\sigma,r}\overline{E}_{p,\sigma,r}$, we need to find
$\overline{E}_{p,\sigma,r}$ as the solution of the Euclidean eigenvalue equation
\bea
   \left[(i\partial_4+eEx_3)^2
    +(i\partial_1)^2+(i\partial_2-eHx_1)^2
    +(i\partial_3)^2-e\left(rE+\sigma H\right)
      \right]\overline{E}_{p,\sigma,r}(x)
    &=&p^{2} \overline{E}_{p,\sigma,r}(x),
\label{epbareucl1}
\eea
Once we performed the procedure shown in
Eqs.~(\ref{euclideanver})-(\ref{epmagneticfull}), we arrive at
\bea
  \overline{\mathbb{E}}_p
   =\overline{\mathbb{E}}_p^{E}(x_{||})\overline{\mathbb{E}}_p^H(x_{\perp})
\label{epbareucl2}
\eea
with
\bea
\overline{\mathbb{E}}_p^{E}(x_{||})
 \equiv {\sum_{r=\pm1}}\overline{E}_{p,r}^{E}(x_\|)
  {\Delta^{E}(r)}=  e^{-ip_4x_4}
  \sum_{r=\pm1}\chi_{p,r}^{E}(x_3)
  {\Delta^{E}(r)}
\label{barepelectricfull}
\eea
\bea
\overline{\mathbb{E}}_p^{H}(x_{\perp})
 \equiv {\sum_{\sigma=\pm1}}\overline{E}_{p,\sigma}^{H}(x_\perp)
  {\Delta^{H}(\sigma)}=
  e^{-ip_2x_2}
  {\sum_{\sigma=\pm1}}\chi_{p,\sigma}^{H}(x_1)
  {\Delta^H(\sigma)}
\label{barepmagneticfull}
\eea

To finish, it is worth to mention that the solutions of (\ref{KGequMink}) in Minkowski space has been worked out
in \cite{Ritus:1978cj, Nikishov, GreinerM&Rafelski}. They are  widely
used in the scattering matrix method. In Minkowski space, the charged
particles in the presence of a uniform electric field do not have
bound states. However, after the Wick rotation to Euclidean variables,
the dispersion by a barrier problem is transformed into that of a
particle in a potential well. The solutions to the field equation in
this case are instanton-type, that is, classical solutions in
Euclidean time~\cite{ashokdas,KimPage}.

\subsection{Properties of the Euclidean $\mathbb{E}_p$ functions}

Starting from (\ref{epfinal}) and (\ref{epbareucl2}), it is easy to check the orthogonality and completeness relations of
the $\mathbb{E}_p$  functions.
One can explore the orthogonality by separating the integration in parallel and transverse coordinates
\bea
   \int d^4x\ \overline{\mathbb{E}}_p(x) \mathbb{E}_{p'}(x)
  =\int d^2x_{||}\
   \overline{\mathbb{E}}_p^{E}(x_{||})\mathbb{E}_{p'}^{E}(x_{||})
   \int d^2x_{\perp}\ \overline{\mathbb{E}}_p^H(x_\perp) {\mathbb{E}_{p'}^H}(x_\perp)
\label{orthosep}
\eea
Let us focus our attention on one of the integrals in the RHS of \Eq{orthosep}. Let's say the one associated with the
electric field.
\bea
  \int d^2x_{||}\
   \overline{\mathbb{E}}_p^{E}(x_{||})
   \mathbb{E}_{p'}^{E}(x_{||})
  &=&
  \sum_{r,r'=\pm1}N_{n_E}N_{{n_E}'}
{  \Delta^{E}(r)\Delta^{E}(r')}
  \int dx_4 dx_3 e^{i({p_4}'-p_4)x_4}D_{n_E}(\rho_{E})D_{{n'_E}}(\rho'_{E})
\eea
where $\rho_{E}\equiv\sqrt{2|eE|}(x_3+p_4/eE)$,  $\rho'_{E}
\equiv\sqrt{2|eE|}(x_3+{p_4}'/eE)$. The integration in $x_4$ produces
a delta function $\delta({p_4}'-p_4)$, which implies that
$\rho_{E}=\rho'_{E}$.

Changing variables from $x_3$ to $\rho_{E}$ and using the orthogonality of the parabolic cylinder functions
\bea
  \int d\rho_{E} D_{n_E}(\rho_{E})D_{n_E'} (\rho_{E})=\sqrt{2\pi}{n_E}!\delta_{{n_E}{n'_E}}
\label{orthoparcyl}
\eea
we obtain
\bea
\int d^2x_{||}\
   \overline{\mathbb{E}}_p^{E}(x_{||})
   \mathbb{E}_{p'}^{E}(x_{||})  &=&
(2\pi)\delta({p_4}'-p_4)
    \sum_{r,r'=\pm1}
N_{n_E}N_{{n_E}'}
    \delta_{rr'}\frac{\sqrt{2\pi}}{\sqrt{2|eE|}}{n_E}!\delta_{{n_E}{n'_E}}
{
   \Delta^{E}(r)}
\nonumber \\
   &=&
(2\pi)^2\delta_{\tilde{l}\tilde{l}'}\delta({p_4}'-p_4)
      \Pi^{E}(\tilde{l})
\label{orthoelectric}
\eea
{where
  $\Pi^{E}(\tilde{l})=\Delta^{E}(sgn(eE))+\Delta^{E}(-sgn(eE))(1-\delta_{\tilde{l}0})$ takes into account the condition $sgn(eE)r=1$ if $\tilde{l}=0$.}
Note that a similar equation is obtained in the external magnetic field case. This similarity  between the magnetic and the
electric fields, is a consequence of the duality between the electric
and the magnetic fields which is evident from the equations of motion
in Euclidean space.

A similar procedure can be carried out for the magnetic part in
\Eq{orthosep}. Thus, the $\mathbb{E}_p$ functions satisfy
the orthogonality condition given by
\bea
   \int d^4x\ \overline{\mathbb{E}}_p(x) \mathbb{E}_{p'}(x)
   =(2\pi)^4\delta_{\tilde{l}\tilde{l}'}\delta_{ll'}
     \delta({p_4}'-p_4)\delta(p'_2-p_2)\Pi^{E}(\tilde{l})\Pi^H(l)
\label{orthonormalityfull}
\eea
where the projectors $\Pi^{E}(\tilde{l})$ and $\Pi^H(l)$ guarantee that in the quantum states with $l=0$ and/or $\tilde{l}=0$, respectively, the particle can have only one spin and/or dipole moment projection.

The completeness relation can be easily proved by using the
orthogonality condition of the $\mathbb{E}_p$ functions, as follows.
Let us multiply  the RHS of \Eq{orthonormalityfull} by
$\overline{\mathbb{E}}_{p'}(y)$ and then integrate over $p'$, 
\bea
    \int d^4x\ \overline{\mathbb{E}}_p(x)
    \sumint \frac{d^4p'}{(2\pi)^4}
    \mathbb{E}_{p'}(x)\overline{\mathbb{E}}_{p'}(y)
   &=&\sumint d^4p'\delta_{\tilde{l}\tilde{l}'}\delta_{ll'}
     \delta({p_4}'-p_4)\delta(p'_2-p_2)\Pi^{E}(\tilde{l})\Pi^H(l)
     \overline{\mathbb{E}}_{p'}(y)
\nonumber \\
   &=&\Pi^{E}(\tilde{l})\Pi^H(l)\overline{\mathbb{E}}_{p}(y)
    =\overline{\mathbb{E}}_{p}(y).
\label{completeness1}
\eea
Thus, from the above equation, we can see that the $\mathbb{E}_p$
functions satisfy
\bea
   \sumint \frac{d^4p'}{(2\pi)^4}
    \mathbb{E}_{p'}(x)\overline{\mathbb{E}}_{p'}(y)=\delta^4(x-y),
\label{completeness2}
\eea
which is the completeness relation of the $\mathbb{E}_p$ functions.

From the orthogonality condition of the $\mathbb{E}_p$ functions we have that the condensate does not depend on the
representation,
\bea
   \int d^4x\ \overline{\psi}(x)\psi(x)
  =\sumint \frac{d^4p}{(2\pi)^4}\overline{\psi}(p)\psi(p)
\label{condensate}
\eea
with $\sumint \frac{d^4p}{(2\pi)^4}= \frac{1}{(2\pi)^4}\sum_{l,\widetilde{l}=0}^{\infty} \int_{-\infty}^{\infty}dp_4 dp_2$.

Relation (\ref{condensate}) follows automatically from the $\mathbb{E}_p$-transformation of the wave functions of charged particles
\bea
  \psi(x)=\sumint\frac{d^4p}{(2\pi)^4}\mathbb{E}_p(x)\psi(p),
\ \ \ \
  \overline{\psi}(x)
  =\sumint\frac{d^4p}{(2\pi)^4}\overline{\psi}(p)\overline{\mathbb{E}}_p(x)
\label{psimomentum}
\eea
together with the orthogonality condition (\ref{orthonormalityfull}).

\subsection{Generalized Momentum for the Parallel-Field Configuration}

To find the generalize momentum $\overline{p}_\mu$ of the charged particles under a parallel constant electromagnetic field we should solve the equation
\bea
    \slsh{\Pi}\mathbb{E}_p=\mathbb{E}_p\gamma\cdot\overline{p}
\label{mainproperty}
\eea

Taking into account Eqs.~(\ref{epsep})-(\ref{epmagneticfull}), we have that the particle dynamics along transverse and longitudinal coordinates are decoupled. Then, the LHS of Eq. (\ref{mainproperty}) can be written as
\bea
    \slsh{\Pi}\mathbb{E}_p=(\slsh{\Pi}^{||}+\slsh{\Pi}^{\perp}){\mathbb{E}}_p^{E}\mathbb{E}_p^H=
    (\slsh{\Pi}^{||}{\mathbb{E}}_p^{E})\mathbb{E}_p^H+{\mathbb{E}}_p^{E}(\slsh{\Pi}^{\perp}\mathbb{E}_p^H)
\label{mainproperty-2}
\eea
where we took into account (\ref{epsep}), and that $\Pi_\mu^{||}=(-i\partial_4+eEx_3,0,0,-i\partial_3)$, $\Pi_\mu^\perp=(0,-i\partial_1,-i\partial_2-eHx_1,0)$.

Now we can separately solve the equations
\bea
   \slsh{\Pi}^{||}{\mathbb{E}}_p^{E}
   &=&{\mathbb{E}_p}^{E}(\gamma^{||}\cdot \overline{p}_{||})
\label{mainelectric}\\
  \slsh{\Pi}^{\perp}\mathbb{E}_p^H
   &=&\mathbb{E}_p^H(\gamma^\perp\cdot\overline{p}_\perp)
\label{mainmagnetic}
\eea
Let us focus our attention on the LHS of \Eq{mainelectric}. Without lost of generality, let us assume that
$sgn(eE)>0 $. Thus, using that $\gamma_3=-i\Sigma_3\gamma_5\gamma_4$, \Eq{mainelectric} can be written as
\bea
   \slsh{\Pi}^{||}{\mathbb{E}}_p^{E}&=&
   e^{ip_4x_4}
\left\{
   \left[\hat{a}\Delta^E(-)+\hat{a}^\dagger\Delta^E(+)\right]\gamma^4
\right\}
     \sum_{r=\pm1}\chi_{p,r}^{E}(x_3)
  \Delta^{E}(r)
\label{mainelecexpand}
\eea
where $\widehat{a}\equiv(p_4+eEx_3)+\partial_3$ and
$\widehat{a}^\dagger\equiv(p_4+eEx_3)-\partial_3$ are lower and
raiser ladder operators, respectively.

Taking into account that $\Delta^E(r)\gamma_4=\gamma_4\Delta^E(-r)$
together with
\begin{equation}\label{a-plus}
\widehat{a}^\dagger \chi_{p,-}^E
=\sqrt{2|eE|\tilde{l}}\ \chi_{p,+}^E
\end{equation}

\begin{equation}\label{a}
\widehat{a} \chi_{p,+}^E
=\sqrt{2|eE|\tilde{l}}\ \chi_{p,-}^E\ ,
\end{equation}
which are easily obtained by using
$\hat{a}D_{\tilde{l}}(\eta)=\tilde{l}\sqrt{2|eE|}D_{\tilde{l}-1}(\eta)$ and
$\hat{a}^\dagger D_{\tilde{l-1}}(\eta)=\sqrt{2|eE|}D_{\tilde{l}}(\eta)$,
we  can rewrite Eq. (\ref{mainelecexpand}) as
\bea
   \slsh{\Pi}^{||}{\mathbb{E}}_p^{E}=
   e^{ip_4x_4}
   \left[\Delta^E(-)\sqrt{2|eE|\tilde{l}}\ \chi_{p,-}^E+\Delta^E(+)\sqrt{2|eE|\tilde{l}}\ \chi_{p,+}^E\right]\gamma^4={\mathbb{E}}_p^{E}(\gamma^{||}\cdot \overline{p}_{||})
\label{maineleexpand2}
\eea
with
\bea
   \overline{p}^{||}_{\mu}=(\sqrt{2\tilde{l}|eE|},0,0,0).
\label{parallelp}
\eea
where we used the fact that for $\tilde{l}\geq 1$, $\Pi^E(\tilde{l})$ commutes with $\gamma_4$, and, when $\tilde{l}=0$ the RHS of \Eq{maineleexpand2} is identically zero.

A similar procedure can be followed with \Eq{mainmagnetic} to obtain
\bea
   \overline{p}^{\perp}_{\mu}=(0,0,\sqrt{2l|eH|},0).
\label{perpp}
\eea

Substituting these results back in (\ref{mainproperty-2}) we have

\bea
\slsh{\Pi}\mathbb{E}_p= {\mathbb{E}}_p^{E}\mathbb{E}_p^H(\gamma^{||}\cdot \overline{p}_{||})+{\mathbb{E}}_p^{E}\mathbb{E}_p^H(\gamma^\perp\cdot\overline{p}_\perp)=\mathbb{E}_p\gamma\cdot\overline{p}
\label{mainproperty-3}
\eea
where the generalized momentum, in the general case, is given by

\bea
   \overline{p}_{\mu}=\overline{p}_\mu^{||}+\overline{p}_\mu^\perp=(sgn(eE)\sqrt{2\tilde{l}|eE|},0,sgn(eH)\sqrt{2l|eH|},0).
\label{overlinep}
\eea

\section{Non-perturbative Euler-Heisenberg Lagrangian for electrons in (1+1)-dimensions
  in a weak electric field.}

As we already mentioned, in the presence of a strong
magnetic field charged fermions suffer a dimensional reduction from
(3+1) to (1+1) dimensions. This is  because all fermions are confined to
the LLL. In this section we shall follow an alternative approach to find the Euler-Heisenberg Lagrangian of the studied system starting from an
effective theory of fermions in the reduced (1+1) dimensions under a weak electric field.

Let us start from the Euclidean action (\ref{actioneuclritus})
\bea
   S&=&\sumint \frac{d^4p}{(2\pi)^4}
   \overline{\psi}(p)
   \Pi^{E}(\tilde{l})\Pi^H(l)
   [-\gamma_\mu\overline{p}_\mu-\tilde{\Sigma}(\overline{p})]
   \psi(p)   .
\label{efflag1}
\eea
Because the factor $\Pi^H(l)$ emphasizes that fermions in the LLL have
only two degrees of freedom compared to the four degrees of freedom of those in higher LL's,
let's separate the contribution of the LLL from the rest
\bea
   S&=&S_0+
   \frac{1}{l_El_H}\sumint \frac{d^4p}{(2\pi)^4}
   \overline{\psi}(p)
   \Pi^{E}(\tilde{l})
   [-\gamma_\mu\overline{p}_\mu-\tilde{\Sigma}(\overline{p})]
   \psi(p)
\label{efflag1a}
\eea
where
\bea
  S_0=\frac{1}{l_El_H}\sum_{\tilde{l}=0}^\infty \int\frac{dp_4dp_2}{(2\pi)^4}
   \overline{\psi}(p)
   \Pi^{E}(\tilde{l})\Delta(+)
   [-\gamma_\mu\overline{p}_\mu-\tilde{\Sigma}(\overline{p})]
   \psi(p)
\label{efflag1b}
\eea
is the action of fermions in the LLL with $l_E\equiv1/\sqrt{eE}$
and $l_H\equiv 1/{\sqrt{eH}}$ being the magnetic and electric
characteristic length scales, respectively. Note that $\Pi^E(l)$ plays
the same role as its magnetic counterpart $\Pi^H(l)$  separating the
dynamics of fermions in the Lowest Electric Level from the rest.
However, because we are interested in a scenario in which the electric
field is weak, we keep all electric levels in $S_0$.

Using the projectors of \Eq{projectors}, the spinor field can be
decomposed as follows
\bea
  \psi^{(+)}_R=\Delta^E(+)\Delta^H(+)\psi \ , \ \
  \psi^{(+)}_L=\Delta^E(-)\Delta^H(+)\psi \ , \ \
  \psi^{(-)}_R=\Delta^E(-)\Delta^H(-)\psi \ , \ \
  \psi^{(-)}_L=\Delta^E(+)\Delta^H(-)\psi \ ,
\label{efflag2} \\
  \overline{\psi}^{(+)}_R=\overline{\psi}\Delta^E(-)\Delta^H(+) \ , \ \
  \overline{\psi}^{(+)}_L=\overline{\psi}\Delta^E(+)\Delta^H(+) \ , \ \
  \overline{\psi}^{(-)}_R=\overline{\psi}\Delta^E(+)\Delta^H(-) \ , \ \
  \overline{\psi}^{(-)}_L=\overline{\psi}\Delta^E(-)\Delta^H(-) \ .
\label{efflag3}
\eea
where the supraindices $(\pm)$ denote the spin-up (+) and spin-down (-) projections, while the subindices (R/L) are labeling the right (R) and left (L) chirality projections.

In (\ref{efflag2}) and (\ref{efflag3}) it was used that $\Delta^H(\sigma)\Delta^E(r)=\Delta^H(\sigma)P(r\sigma)$ with
\bea
  P(\sigma r)= \frac{1}{2}\left(1+r\sigma\gamma_5\right),
\label{chiralprojector}
\eea
the usual chiral projector: $\psi_R=P(+)\psi $ and $\psi_L=P(-)\psi$.

It can be easily checked that
\bea
  \psi=\psi^{(+)}_{R}+\psi^{(+)}_{L}+\psi^{(-)}_{R}+\psi^{(-)}_{L}.
\label{efflag4}
\eea

Then, using \Eq{efflag2} and \Eq{efflag3}, we rewrite the effective
action for the LLL of (\ref{efflag1b}) as
\bea
  S_0&=&\frac{1}{l_El_H}\sum_{\tilde{l}=0}^\infty\int\frac{dp_4dp_2}{(2\pi)^4}
   \left\{
  -\overline{\psi}^{(+)}_R(p)
   \Pi^{E}(\tilde{l})\gamma_\mu\overline{p}_\mu
   \psi^{(+)}_{R}(p)
  -\overline{\psi}^{(+)}_L(p)
   \Pi^{E}(\tilde{l})\gamma_\mu\overline{p}_\mu
   \psi^{(+)}_{L}(p)
\right.
\nonumber \\
&&\hspace{+2.5cm}-
\left.
  \mathcal{E}^0
   \left[
   \overline{\psi}^{(+)}_R(p)
   \Pi^{E}(\tilde{l})
   \psi^{(+)}_{L}(p)
  +\overline{\psi}^{(+)}_L(p)
   \Pi^{E}(\tilde{l})
   \psi^{(+)}_{R}(p)
    \right]
   \right\}.
\label{effelag5}
\eea
where $\Sigma(\overline{p})=\Delta(+)\mathcal{E}^0$.

Writing the four Dirac spinor in the LLL as
$\psi^T=(\psi_1,\psi_2,\psi_3,\psi_4)$, \Eq{effelag5} reduces to
\bea
   S_0=\frac{1}{l_El_H}\sum_{\tilde{l}=0}^\infty\int\frac{dp_4dp_2}{(2\pi)^4}
   \left(i\psi_3^*,i\psi_1^*\right)
\tilde{\Pi}^E(\tilde{l})
   \left(
   \begin{array}{cc}
   -\mathcal{E}^0 & -i \sqrt{2|eE|\tilde{l}} \\
    -i \sqrt{2|eE|\tilde{l}}        & -\mathcal{E}^0
   \end{array}
   \right)
   \left(
   \begin{array}{c}
    \psi_1 \\
    \psi_3
   \end{array}\right)
\label{effelag6}
\eea
where
$\tilde{\Pi}(\tilde{l})=\frac{1}{2}(1+\sigma_3\delta_{\tilde{l}0})$
with $\sigma_3$ the Pauli's matrix.

Introducing the Dirac gamma matrices in the (1+1) Euclidean space
\bea
  \tilde{\gamma}_4=i\sigma_1 \ , \hspace{1cm} \tilde{\gamma}_1=-i\sigma_2
   \hspace{1cm}\mbox{and}\hspace{1cm}
  \tilde{\gamma}_5=\sigma_3=i\tilde{\gamma_1}\tilde{\gamma_4}\ ,
\label{efflag7}
\eea
which satisfy
$\{\tilde{\gamma}_\mu,\tilde{\gamma}_\nu\}=-2\delta_{\mu\nu}$ with $\delta_{\mu\nu}=diag(1,1)$,
and the 1+1 spinor field in the LLL as
\bea
   \psi_{LLL}\equiv
   \left(
   \begin{array}{c}
    \psi_1 \\
    \psi_3
   \end{array}\right),
\label{efflag8}
\eea
then we rewrite \Eq{effelag6} as
\bea
   S_0=\frac{1}{l_El_H}\sum_{\tilde{l}=0}^\infty\int\frac{dp_4dp_2}{(2\pi)^4}
   \overline{\psi}_{LLL}\tilde{\Pi}^E(\tilde{l})
   \left(
   -\widetilde{\gamma}_\mu\widetilde{p}_\mu-\mathcal{E}^0
   \right)
   \psi_{LLL}
\label{efflag9}
\eea
where $\widetilde{p}_\mu=(\sqrt{2|eE|\tilde{l}},0)$.

Note that the above equation in the limit $E\rightarrow0$ describes a
theory of fermions in the LLL of the $MC\chi SB$ phase \cite{ferrerincera}. This statement becomes evident once we make the replacement $(1/l_E)\sum_{\widetilde{l}} \rightarrow\int dp_3$, and
$\widetilde{p}_\mu \rightarrow \tilde{p}^\|_\mu=(p_4,p_3)$.

Taking into account \Eq{efflag9}, the Euclidean 1-loop effective
action $\Phi^1_E$ for electrons in the LLL has the form
\bea
   e^{\Phi^1_E}&=&\int [\mathcal{D}\{\overline{\psi}_{LLL}\}]
   [\mathcal{D}\{\psi_{LLL}\}] e^{S_0}
\nonumber \\
  &=&\mathcal{D}et\left[\tilde{\Pi}^E(\tilde{l})
     (-\widetilde{\gamma}_\mu\widetilde{p}_\mu-\mathcal{E}^0)\right].
\label{effeactb14}
\eea
We can rewrite the effective action in \Eq{effeactb14} as
\bea
   \phi^1_E&=&\frac{1}{l_El_H}\int\frac{dp_4dp_2}{(2\pi)^2}
    \sum_{\tilde{l}=0}^\infty (2-\delta_{0\tilde{l}})\frac{1}{2}
    \ln\left[\overline{p}^2+(\mathcal{E}^0)^2\right]
\label{1loopactionLLL}
\eea
where $\phi^1_E\equiv(l_El_H)^{-1}\Phi^1_E$ and we took into account the
definition of $\Pi^E(\tilde{l})$. Once we integrate out
$p_4$ and $p_2$, we get
\bea
   \phi^1_E&=&\frac{\left|eH\right|\left|eE\right|}{2(2\pi)^2}
    \sum_{\tilde{l}=0}^\infty (2-\delta_{0\tilde{l}})
    \int_{l_H^{2}}^\infty
    \frac{ds}{s}e^{-s[\overline{p}^2+(\mathcal{E}^0)^2]}
\nonumber \\
&=&\frac{|eH|}{2(2\pi)^2}
    \int_{l_H^{2}}^\infty
    \frac{ds}{s^2}e^{-s(\mathcal{E}^0)^2}(|eE|s) \coth(eEs)
\label{1loopactionLLLsum}
\eea
where we used the ultraviolet cutoff  $l_H^{2}$ in $s$ that accounts for the LLL  dominance of the pairing dynamics in this theory. Note that
\Eq{1loopactionLLLsum} coincides with \Eq{w1HstrongEweak} as
expected.

\end{document}